\documentclass[pre,twocolumn,floatfix]{revtex4}

\usepackage{graphicx}
\usepackage{bm}
\usepackage{url}

\begin{document}

\title{Statistical mechanics of money, debt, and energy consumption}

\author{Victor M.~Yakovenko}

\affiliation{Department of Physics, University of
  Maryland, College Park, Maryland 20742-4111, USA}

\date{12 August 2010}

\begin{abstract}
We briefly review statistical models for the probability distribution of money developed in the econophysics literature since the late 1990s.  In these models, economic transactions are modeled as random transfers of money between the agents in payment for goods and services.  We focus on conceptual foundations for this approach, on the issues of money conservation and debt, and present new results for the energy consumption distribution around the world. \\

\textsf{\normalsize ``Money, it's a gas.'' Pink Floyd, 
\textit{Dark Side of the Moon}}
\end{abstract}

\maketitle

\section{Introduction}
\label{Sec:money}

Econophysics is an interdisciplinary field applying mathematical methods of statistical physics to social, economical, and financial problems \cite{Stauffer-history}.  The term was first introduced by the statistical physicist Eugene Stanley at the conference \textit{Dynamics of Complex Systems} in Kolkata in 1995 \cite{Chakrabarti-history} and printed in its proceedings  \cite{Stanley-1996}.

A puzzling social problem is the persistent economic inequality among the population in any society.  In statistical physics, it is very well known that identical (``equal'') molecules in a gas spontaneously develop a widely unequal distribution of energies as a result of random energy transfers in molecular collisions.  By analogy, a very unequal probability distribution of money can develop spontaneously as a result of random money transfers between economic agents.  This idea was proposed by several econophysicists around 2000 \cite{Ispolatov-1998,Dragulescu-2000,Chakraborti-2000,Mimkes-2000} and much earlier by the sociologist John Angle \cite{Angle-1986}.  The subsequent progress is reviewed in \cite{Yakovenko-2009,Chatterjee-2007,Richmond-2006a,Richmond-2006b}
and popular articles \cite{Hayes-2002,Hogan-2005,Shea-2005}.  This novel approach has virtually no counterpart in the economic literature.  Only the economist Miguel Molico \cite{Molico-2006} recently studied the probability distribution of money within the search theory of money \cite{Kiyotaki-1993}.  Econophysics ideas are gradually starting to receive recognition from the economists and social scientists \cite{conferences}. 

Econophysics papers typically focus on calculations and analysis of intricate details of mathematical models, but not on conceptual foundations.  This shortcoming was criticized by the economists \cite{Gallegati-2006}.  In this article, we present an extended discussion of the conceptual foundation underlying the models of random money transfers.   We focus on the often-contentious issues of money conservation and debt, and also present new results on the energy consumption distribution around the world.  For a more comprehensive survey of the literature, please refer to the review papers cited above.

\section{The Boltzmann-Gibbs distribution of energy}
\label{Sec:BGphysics}

The fundamental law of equilibrium statistical physics is the
Boltzmann-Gibbs distribution of energy \cite{Wannier-book}.  It states that the probability $P(\varepsilon)$ of finding a physical system or subsystem in a state
with the energy $\varepsilon$ is given by the exponential function
\begin{equation}
  P(\varepsilon)=c\,e^{-\varepsilon/T}.
\label{Gibbs}
\end{equation}
Here $c$ is a normalizing constant, and $T$ is the temperature, which is equal to the average energy per particle: $T\sim\langle\varepsilon\rangle$, up to a coefficient of the order of 1.

A derivation of Eq.\ (\ref{Gibbs}) involves the two main
ingredients: statistical character of the system and conservation of
energy $\varepsilon$ \cite{Wannier-book}.  One of the shortest derivations can be
summarized as follows.  Let us divide the system into two (generally
unequal) parts.  Then, the total energy is the sum of the parts:
$\varepsilon=\varepsilon_1+\varepsilon_2$, whereas the probability is
the product of probabilities:
$P(\varepsilon)=P(\varepsilon_1)\,P(\varepsilon_2)$.  The only
solution of these two equations is the exponential function
(\ref{Gibbs}).  Eq.\ (\ref{Gibbs}) can be also derived by maximizing the entropy $S=-\sum_k N_k\ln(N_k/N)$ of the system for a fixed total energy $E=\sum_kN_k\varepsilon_k$, where $N_k$ is the number of particles having the energy $\varepsilon_k$.

These derivations are very general, so one may expect that the exponential
distribution (\ref{Gibbs}) would apply to other statistical systems with a conserved quantity.

\section{Conservation of money}
\label{Sec:conservation}

The economy is a big statistical system with millions of participating
agents, so it is a promising target for applications of statistical
mechanics.  Is there a conserved quantity in the economy?
Dr\u{a}gulescu and Yakovenko \cite{Dragulescu-2000} argued that such a conserved quantity is money $m$.  Indeed, the ordinary economic agents can only receive
money from and give money to other agents.  They are not permitted to
``manufacture'' money, e.g.,\ to print dollar bills.  Let us consider
an economic transaction between agents $i$ and $j$.  When the agent
$i$ pays money $\Delta m$ to the agent $j$ for some goods or services,
the money balances of the agents change as follows
\begin{eqnarray}
  && m_i\;\rightarrow\; m_i'=m_i-\Delta m,
\nonumber \\
  && m_j\;\rightarrow\; m_j'=m_j+\Delta m.
\label{transfer}
\end{eqnarray}
The total amount of money of the two agents before and after
transaction remains the same
\begin{equation}
  m_i+m_j=m_i'+m_j',
\label{conservation}
\end{equation}
i.e.,\ there is a local conservation law for money.  The transfer of money (\ref{transfer}) is analogous to the transfer of energy in molecular collisions, and Eq.\ (\ref{conservation}) is analogous to conservation of energy.  Conservative models of this kind are also studied in some economic literature \cite{Kiyotaki-1993,Molico-2006}.

We should emphasize that the transfer of money from one agent to another represents payment for goods and services in a market economy.  However, Eq.~(\ref{transfer}) only keeps track of money flow, but does not keep track of what goods and service are delivered.  One reason for this is that many goods, e.g.,\ food and other supplies, and most services, e.g.,\ getting a haircut or going to a movie, are not
tangible and disappear after consumption.  Because they are not
conserved, and also because they are measured in different physical
units, it is not practical to keep track of them.  In contrast,
money is measured in the same unit (within a given country with a
single currency) and is conserved in local transactions
(\ref{conservation}), so it is straightforward to keep track of money
flow.  It is also important to realize that an increase in material
production does not result in an automatic increase in money supply.
The agents can grow apples on trees, but cannot grow money on trees.

Enforcement of the local conservation law (\ref{conservation}) is crucial for successful functioning of money.  If the agents were
permitted to ``manufacture'' money, they would be printing money and
buying all goods for nothing, which would be a disaster.  The purpose of the conservation law is to ensure that an agent can buy goods from the society only if he or she contributes something useful to the society and receives monetary payment for these contributions.  Money is an accounting device, and, indeed, all accounting systems are based on the conservation law (\ref{transfer}).  The physical medium of money is not essential as long as the
conservation law is enforced.  The fiat money (declared to be money by
the central bank) may be in the form of paper currency, but more often it is represented by digits on computer accounts.  So, money is just bits of information, and monetary system constitutes an informational layer of the economy.  Monetary system interacts with physical system (production and consumption of material goods), but the two layers cannot be transformed into each other because of their different nature.

Unlike, ordinary economic agents, a central bank or a central
government can inject money into the economy, thus changing the total
amount of money in the system.  This process is analogous to an influx
of energy into a system from external sources.  As long as the rate of money influx is slow compared with the relaxation rate in the economy, the system remains in a quasi-stationary statistical equilibrium with slowly changing parameters.  This situation is analogous to slow heating of a kettle, where the kettle
has a well defined, but slowly increasing, temperature at any moment of
time.

Another potential problem with conservation of money is debt, which will be discussed in Sec.~\ref{Sec:debt}.  Most of econophysics models, such as \cite{Ispolatov-1998,Dragulescu-2000,Chakraborti-2000}, and some economic models \cite{Kiyotaki-1993,Molico-2006} do not permit debt.  Thus, money balances of the agents cannot drop below zero: $m_i\geq0$ for all $i$.  Transaction (\ref{transfer}) takes place only when an agent has enough money to pay the price: $m_i\geq\Delta m$.  An agent with $m_i=0$ cannot buy goods from other agents, but can receive money for delivering goods or services to them.

In a big statistical ensemble of agents, monetary transactions (\ref{transfer}) take place between many different agents.  Although purposeful and rational for individual agents, these transactions can be treated as effectively random for the whole ensemble.  This is similar to statistical physics, where each atom follows deterministic equations of motion, but the whole system is effectively random.

We are interested in the probability distribution of money $P(m)$ among the economic agents resulting from the random transfers (\ref{transfer}).  For this purpose, it is appropriate to make the simplifying macroeconomic idealizations, as described above, in order to ensure overall stability of the system and existence of statistical equilibrium.  The concept of ``equilibrium''
is a very common idealization in economic literature, even though the
real economy might never be in equilibrium.  We extend this
concept to a statistical equilibrium, characterized by a
stationary probability distribution $P(m)$, in contrast to a
mechanical equilibrium, where the ``forces'' of demand and supply
balance each other.

\section{Probability distribution of money}
\label{Sec:BGmoney}

Having recognized the principle of local money conservation,
Dr\u{a}gulescu and Yakovenko \cite{Dragulescu-2000} argued that the  distribution of money $P(m)$ should be given by the exponential Boltzmann-Gibbs
function analogous to Eq.\ (\ref{Gibbs})
\begin{equation}
  P(m)=c\,e^{-m/T_m}.
\label{money}
\end{equation}
Here $c$ is a normalizing constant, and $T_m$ is the ``money
temperature'', which is equal to the average amount of money per
agent: $T=\langle m\rangle=M/N$, where $M$ is the total money, and $N$
is the number of agents.

\begin{figure}
\includegraphics[width=0.9\linewidth]{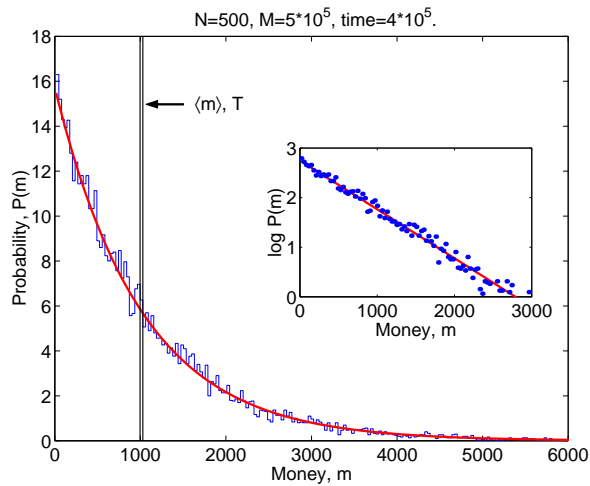}
\caption{\textit{Histogram and points:} Stationary probability
  distribution of money $P(m)$ obtained in the random transfers model
  \cite{Dragulescu-2000}.  \textit{Solid curves:} Fits to the exponential
  distribution (\ref{money}).  \textit{Vertical line:} The initial distribution of
  money.}
\label{Fig:money}
\end{figure}

To verify this conjecture, Dr\u{a}gulescu and Yakovenko \cite{Dragulescu-2000} performed agent-based computer simulations of money transfers between agents.
Initially all agents were given the same amount of money, say, \$1000.
Then, a pair of agents $(i,j)$ was randomly selected, the amount
$\Delta m$ was transferred from one agent to another, and the process
was repeated many times.  Time evolution of the probability
distribution of money $P(m)$ is shown in computer animation videos \cite{Chen-2007} and \cite{Wright-2007}.  After a transitory period, money distribution converges to the stationary form shown in Fig.\ \ref{Fig:money}.  As expected, the distribution is well fitted by the exponential function (\ref{money}).

In the simplest model \cite{Dragulescu-2000}, the transferred amount was
fixed to a constant $\Delta m=\$1$.  Computer animation \cite{Chen-2007} shows that the initial distribution of money first broadens to a symmetric Gaussian curve,
typical for a diffusion process.  Then, the distribution starts
to pile up around the $m=0$ state, which acts as the impenetrable
boundary, because of the condition $m\geq0$.  As a result,
$P(m)$ becomes skewed (asymmetric) and eventually reaches the
stationary exponential shape, as shown in Fig.~\ref{Fig:money}.  The
boundary at $m=0$ is analogous to the ground-state energy in
statistical physics.  Without this boundary condition, the probability
distribution of money would not reach a stationary state.  Computer
animations \cite{Chen-2007,Wright-2007} also show how the entropy of
money distribution, defined as $S/N=-\sum_k P(m_k)\ln P(m_k)$, grows
from the initial value $S=0$, where all agents have the same money, to
the maximal value at the statistical equilibrium.

Dr\u{a}gulescu and Yakovenko \cite{Dragulescu-2000} studied different additive rules for $\Delta m$.  Other papers studied multiplicative rules, such as the proportional rule $\Delta m=\gamma m_i$ \cite{Ispolatov-1998,Angle-1986}, the saving propensity \cite{Chakraborti-2000}, and negotiable price \cite{Molico-2006}.  These models produce Gamma-like distributions, as well as a power-law tail for a random distribution of saving propensities.  Despite some mathematical differences, all these models demonstrate spontaneous development of a highly unequal probability distribution of money as a result of random money transfers between the agents.  Many papers use the term ``wealth'' instead of ``money''.  We believe that these terms have different meanings and should not be used interchangeably \cite{Dragulescu-2000,Yakovenko-2009}.

It would be very interesting to compare these theoretical results with empirical data on money distribution.  Unfortunately, it is very difficult to obtain such data.  The probability distribution of balances on deposit accounts in a big enough bank would be a reasonable approximation for money distribution among the population. However, such data are not publicly available.  In contrast, plenty of data are available on income distribution from the tax agencies.  Quantitative analysis of such data for the USA \cite{Yakovenko-2009} shows that the population consists of two distinct social classes.  Income distribution follows the exponential law for the lower class (about 97\% of population) and the power law for the upper class (about 3\% of population).   Although social classes have been known since Karl Marx, it is interesting that they can be recognized by fitting the empirical data with simple mathematical functions.  The computer scientist Ian Wright \cite{Wright-2005,Wright-2009} has demonstrated emergence of two classes in sophisticated agent-based simulations of initially equal agents.  This work is further developed in the book \cite{Cockshott-2009}, integrating economics, computer science, and physics.

\section{Models of debt}
\label{Sec:debt}

Now let us discuss how the results change when debt is
permitted.  From the standpoint of individual
economic agents, debt may be considered as negative money.  When an
agent borrows money from a bank \cite{bank}, the cash balance of the agent (positive
money) increases, but the agent also acquires a debt obligation
(negative money), so the total balance (net worth) of the agent
remains the same.  Thus, the act of money borrowing still satisfies a
generalized conservation law of the total money (net worth), which is
now defined as the algebraic sum of positive (cash $M$) and negative
(debt $D$) contributions: $M-D=M_b$, where $M_b$ is the original amount of money in the system, the monetary base \cite{McConnell-book}.  When an agent needs to buy a product at a price $\Delta m$ exceeding his money balance $m_i$, the agent is now permitted to borrow the difference from a bank.  After the transaction, the new balance of the agent becomes negative: $m_i'=m_i-\Delta m<0$.  The local conservation law (\ref{transfer}) and (\ref{conservation}) is still satisfied, but now it involves negative values of $m$.  Thus, the consequence of debt is not a violation of the conservation law, but a modification of the boundary condition by permitting negative balances $m_i<0$, so $m=0$ is not the ground state any more.

If the computer simulation with $\Delta m=\$1$ is repeated without any restrictions on the debt of the agents, the probability distribution of money $P(m)$ never stabilizes, and the system never reaches a stationary state.  As time goes on, $P(m)$ keeps spreading in a Gaussian manner unlimitedly toward $m=+\infty$
and $m=-\infty$.  Because of the generalized conservation law, the first moment of the algebraically defined money $m$ remains constant $\langle m\rangle=M_b/N$.  It means that some agents become richer with positive balances $m>0$ at the expense of
other agents going further into debt with negative balances $m<0$.

Common sense, as well as the experience with the current financial
crisis, tell us that an economic system cannot be stable if unlimited
debt is permitted \cite{Soddy-book}.  In this case, the agents can
buy goods without producing anything in exchange by simply going
into unlimited debt.  Arguably, the current financial crisis is caused the enormous debt accumulation in the system, triggered by subprime mortgages and financial derivatives based on them. 

Detailed discussion of the current economic situation is not a subject
of this paper.  Going back to the idealized model of money transfers,
one would need to impose a modified boundary condition in
order to prevent unlimited growth of debt and to ensure overall
stability of the system.  Dr\u{a}gulescu and Yakovenko \cite{Dragulescu-2000} considered a simple model where the maximal debt of each agent is limited to $m_d$.  
In this model, $P(m)$ again has the exponential shape, but with the new boundary condition at $m=-m_d$ and the higher money temperature $T_d=m_d+M_b/N$.  By allowing agents to go into debt up to $m_d$, we increase the amount of money available to each agent by $m_d$.

\begin{figure}
\includegraphics[width=\linewidth]{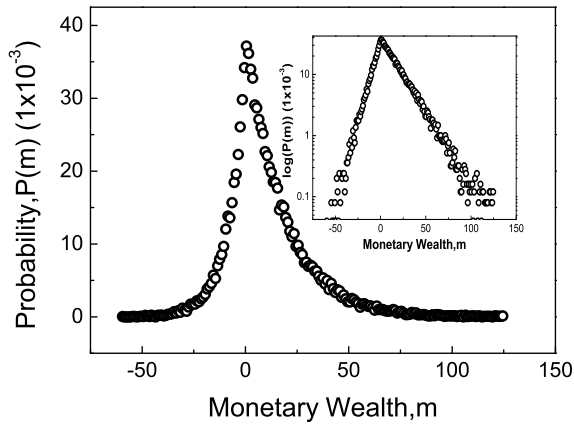}
\caption{The stationary distribution of money \cite{Xi-Ding-Wang-2005} for the 
  required reserve ratio $R=0.8$.  The distribution is exponential for both 
  positive and negative money with different ``temperatures'' $T_+$ and $T_-$, as
  shown in the inset on log-linear scale.}
\label{Fig:reserve}
\end{figure}

Xi, Ding, and Wang \cite{Xi-Ding-Wang-2005} considered a more realistic
boundary condition, where a constraint is imposed on the total debt of all agents in the system.  This is accomplished via the required reserve ratio $R$ \cite{McConnell-book}.  Banks are required by law to set aside a fraction $R$ of the money deposited into bank accounts, whereas the remaining fraction $1-R$ can be lent.  If the initial amount of money in the system (the money
base) is $M_b$, then, with repeated lending and borrowing, the total
amount of positive money available to the agents increases to
$M=M_b/R$, where the factor $1/R$ is called the money multiplier
\cite{McConnell-book}.  This is how ``banks create money''.  This extra money comes from the increase in the total debt $D=M_b/R-M_b$.  Given the two constraints on $M$ and $D$, we expect to find the exponential distributions of positive and negative money characterized by two different temperatures: $T_+=M_b/RN$ and $T_-=M_b(1-R)/RN$.  This is exactly what was found in computer simulations
\cite{Xi-Ding-Wang-2005} shown in Fig.\ \ref{Fig:reserve}.

However, in the real economy, the reserve requirement is not effective in
stabilizing debt in the system, because it applies only to
deposits from general public, but not from corporations.  Moreover, there are alternative instruments of debt, including derivatives and various unregulated ``financial innovations''.  As a result, the total debt is not limited in practice
and can reach catastrophic proportions.  Here we briefly
discuss several models with non-stationary debt.  Dr\u{a}gulescu and Yakovenko \cite{Dragulescu-2000} studied a model with different interest rates for deposits into and loans from a bank.  Computer simulations show that, depending on the choice of parameters, the total amount of money in circulation either increases or decreases in time.  The interest amplifies the destabilizing effect of debt, because positive balances become even more positive and negative even more negative.  A macroeconomic model studied by the economist
Steve Keen \cite{Keen-1995,Keen-2000} exhibits a debt-induced breakdown, where all economic activity stops under the burden of heavy debt and cannot be restarted without a ``debt moratorium''.  The interest rates were fixed in these models
and not adjusted self-consistently.  Cockshott and Cottrell \cite{Cockshott-2008}
proposed a mechanism, where the interest rates are set to cover
probabilistic withdrawals of deposits from a bank.  In an agent-based
simulation, they found that money supply increases first, and then the economy crashes under the weight of accumulated debt.

Bankruptcy is a mechanism for debt stabilization.  It erases the debt of an agent (the negative money) and resets the balance to zero.  However, somebody else (a bank or a lender) counted this debt as a positive asset, which also becomes erased.  In the language of physics, creation of debt is analogous to particle-antiparticle generation (creation of positive and negative money), whereas cancellation of debt corresponds to particle-antiparticle annihilation (annihilation of positive and negative money).  The former and latter dominate during economic booms and busts and represents monetary expansion and contraction.  Bankruptcy is the crucial mechanism for stabilization of money distribution, but it is often overlooked by the economists.  Interest rates are meaningless without a mechanism specifying when bankruptcy is triggered. 

Numerous failed attempts were made to create alternative community money from scratch.  In such a system, when an agent provides goods or services to another agent, their accounts are credited with positive and negative tokens, as in Eq.\ (\ref{transfer}).  However, because the initial global money balance is zero in this case, the probability distribution of money $P(m)$ is symmetric with respect to positive and negative $m$.  Unless a boundary condition is imposed on the lower side, $P(m)$ never stabilizes.  Some agents accumulate unlimited negative balances by consuming goods and services and not contributing anything in return, thus undermining the system.  A capitalist society imposes a lower bound on money balances, whereas a socialist one may consider an upper bound \cite{upper}. 

\section{Probability distribution of energy consumption}
\label{Sec:energy}

\begin{figure}
\includegraphics[width=\linewidth]{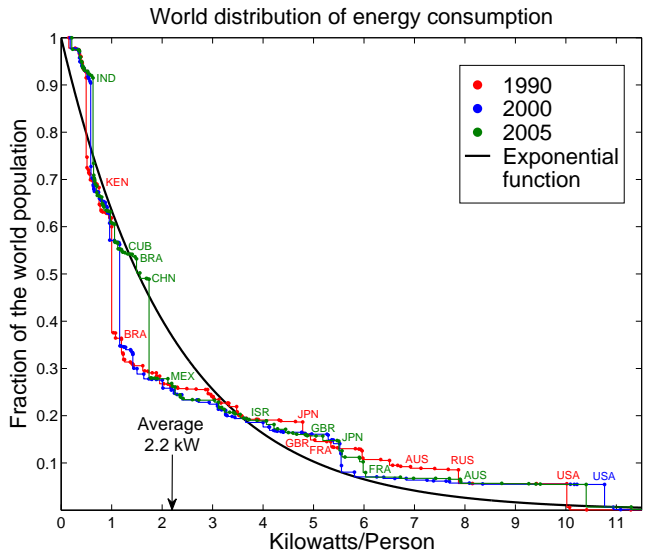}
\caption{Cumulative distribution functions of the energy consumption per capita around the world for 1990, 2000, and 2005.  The solid curve is the exponential function. }
\label{fig:eLinlin}
\end{figure}

While money is the informational side of the economy, material standards of living
are controlled by the physical layer.  They are primarily determined by the level of energy consumption and are widely different around the globe.  Using the data from the World Resources Institute, Banerjee and Yakovenko \cite{Banerjee-2010} found that the probability distribution of energy consumption per capita around the world approximately follows the exponential law, as shown in Fig.~\ref{fig:eLinlin}.  The limited energy resources in the world (predominantly fossil fuels) are partitioned among the world population.  As in Sec.~\ref{Sec:BGphysics}, maximization of the entropy with the constraint results in the exponential distribution of energy consumption.

The world average energy consumption per capita is about 2.2 kW, compared with 10 kW in USA and 0.6 kW in India \cite{Banerjee-2010}.  However, if India and other countries were to adopt the same energy consumption level per capita as in USA, there would not be enough energy resources in the world to do that.  The global energy consumption inequality results from the constraint on energy resources.

\begin{figure}[b]
\includegraphics[width=0.83\linewidth]{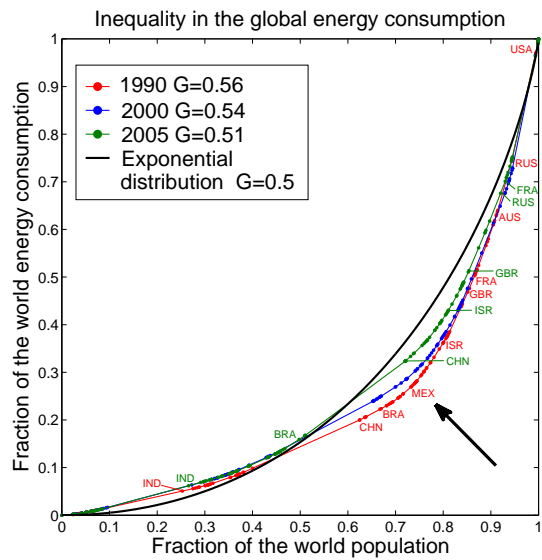}
\caption{Lorenz curves for the energy consumption per capita around the world in 1990, 2000, and 2005, compared with the Lorenz curve for the exponential distribution.}
\label{fig:elorenz}
\end{figure}

Fig.~\ref{fig:elorenz} shows the Lorenz curves for the global energy consumption
in 1990, 2000, and 2005 \cite{Banerjee-2010}.  The horizontal axis in Fig.~\ref{fig:elorenz} represents the cumulative population ranked by the energy consumption per capita, and the vertical axis represents the cumulative energy consumption of this population.  The black solid line is the theoretical curve $y=x+(1-x)\ln(1-x)$ for an exponential distribution \cite{Yakovenko-2009}.  
In the Lorenz plot for 1990, one can notice a kink or a knee indicated by the arrow, where the slope of the curve changes appreciably.  This point represents the boundary between developed and developing countries.  Mexico, Brazil, China, and India are below this point, whereas Britain, France, Japan, Australia, Russia, and USA are above.  Thus, the difference between developed and developing countries lies in the degree of energy consumption and utilization, rather than in the more ephemeral monetary measures, such as dollar income per capita.  However, the Lorenz curve for 2005 is closer to the exponential curve, and the kink is less pronounced.  
It means that the energy consumption inequality and the gap between developed and developing countries have decreased, as also confirmed by the decrease in the Gini coefficient $G$ listed in Fig.~\ref{fig:elorenz}.  We attribute this result to the rapid globalization of the world economy in the last 20 years. 
Ultimately, the energy consumption distribution in a well-mixed globalized world economy is expected to be exponential and not equal.

The energy/ecology and financial/economic crises are the biggest challenges faced by the mankind today.  There is an urgent need to find ways for a manageable and realistic transition from the current breakneck growth-oriented economy, powered by the ever-expanding use of fossil fuels, to a stable and sustainable society, based on renewable energy and balance with the Nature.  Undoubtedly, both money and energy will be the key factors shaping up the future of human civilization.



\begin{thebibliography}{99}
\bibitem{bank} Here we treat the bank as being outside of the system
  consisting of ordinary agents, because we are interested in money
  distribution among these agents.  The debt of agents is an asset for
  the bank, and deposits of cash into the bank are liabilities of the
  bank \cite{McConnell-book}.
\end{thebibliography}
\end{document}